\documentclass[12pt]{article}
\pdfoutput=1
\usepackage{graphicx}

\usepackage{amsfonts}
\usepackage{amsmath}
\usepackage{epsfig}
\usepackage{latexsym}
\usepackage{subfigure}
\usepackage{amssymb}
\numberwithin{equation}{section}
\allowdisplaybreaks

\def\spa#1{\phantom{\fbox{\rule[-#1cm]{0cm}{0cm}}}}

\def\be{\begin{equation}}
\def\ee{\end{equation}}
\def\bea{\begin{eqnarray}}
\def\eea{\end{eqnarray}}

\def\half{{1\over 2}}

\def\del{\partial}
\def\nn{\nonumber}

\renewcommand{\thefootnote}{\fnsymbol{footnote}}

\begin{document}

\hfuzz=100pt
\title{{\Large \bf{Scale invariance and a gravitational model \\ with non-eternal inflation}}}
\date{}
\author{Carlos Herdeiro$^a$\footnote{herdeiro@ua.pt} and Shinji Hirano$^b$\footnote{hirano@eken.phys.nagoya-u.ac.jp}
  \spa{0.5} \\
\\
$^a${{\it Departamento de F\'isica da Universidade de Aveiro and I3N}}
\\ {{\it Campus de Santiago, 3810-183 Aveiro, Portugal}}\\
\\
$^{b}${{\it Department of Physics}}
\\ {{\it Nagoya University}}
\\ {{\it Nagoya 464-8602, Japan}}
  \spa{0.5} 
}
\date{}

\maketitle
\centerline{}

\begin{abstract}
We propose a $3 + 1$ dimensional model of gravity which results in inflation at early times, followed by radiation- and matter-dominated epochs and a subsequent acceleration at late times. Both the inflation and late time acceleration are nearly de Sitter with a large hierarchy between the effective cosmological constants. There is no scalar field agent of inflation, and the transition from the inflation to the radiation-dominated period is smooth. This model is designed so that it yields, at the cost of giving up on Lorentz invariance in the gravitational sector, the Dirac-Born-Infeld type conformal scalar theory when the universe is conformally flat. It, however, resembles Einstein's gravity with the Gibbons-Hawking-York boundary term in weakly curved space-times.
\end{abstract}

\renewcommand{\thefootnote}{\arabic{footnote}}
\setcounter{footnote}{0}

\newpage

\section{Introduction}
Scale invariance is ubiquitous in physics and cosmology is no exception. Observations suggest that the universe is nearly scale invariant at early and late times, when it is believed to be approximately de Sitter space,  with the accelerated expansion being driven by a nearly constant vacuum energy.  

\medskip
The approximate scale invariance at early times can be regarded as evidence for cosmic inflation \cite{Guth:1980zm, Linde:1981mu, Albrecht:1982wi, Starobinsky:1980te, Sato:1980yn} (see, for more recent reviews, \cite{Lyth:1998xn, Linde:2007fr, Rubakov:2010hq, Baumann:2008aq}).
Particularly successful is the slow-roll paradigm \cite{Linde:1981mu, Albrecht:1982wi, Linde:2007fr}. Not only does it solve fundamental problems of the standard Big-Bang cosmology, such as horizon, flatness, and monopole problems, but it also makes important predictions, such as a slight tilt from the scale invariant power spectrum of the cosmic microwave background (CMB) \cite{Komatsu:2010fb}. 
This paradigm typically involves a scalar field, called inflaton, which acts as the agent of the near exponential expansion of the universe. Such an effective theory framework leaves open the lingering question of the particle physics origin of the inflaton. Although fundamental approaches, such as String Theory, have abundant inflaton candidates, (see, for example, \cite{HenryTye:2006uv, Kallosh:2007ig, Burgess:2007pz, McAllister:2007bg}), there is hardly any principle for model selection within many possibilities.
In the absence of a final answer to the fundamental nature of the inflaton, it is worthwhile to explore alternatives to inflaton-driven inflation.

\medskip
In this paper we propose a new model of inflation. Instead of introducing scalar field agents, we alter Einstein's gravity in a manner that automatically results in inflation at early times. Our aim is not to propose the fundamental theory of quantum gravity but rather a simple classical model which might well explain the  observational data. 
Our model construction is, however, guided by scale invariance and thus the model is, to some extent, top-down rather than bottom-up.
Interestingly, not only does our model result in inflation, but it can also accommodate radiation- and matter-dominated epochs and a subsequent late time acceleration with a smooth transition from inflation to radiation. 

\medskip
Our model is designed so that it yields the Dirac-Born-Infeld (DBI) type conformal scalar theory when the universe is conformally flat and resemples Einstein's gravity with the Gibbons-Hawking-York boundary term in weakly curved space-times. An uneasy but important feature is that it breaks down Lorentz invariance by introducing a preferred time-like vector. This is reminiscent of Einstein-aether theory \cite{Jacobson:2008aj}; but the time-like vector is not an independent degree of freedom from the metric, in our model.
Such breakdown occurs in the gravitational sector, where observational constraints are less stringent than in the matter sector. Moreover, Lorentz invariance is recovered in the weak curvature limit. Our setup is also similar to Ho\v{r}ava-Lifshitz gravity \cite{Horava:2009uw} in the sense that the breakdown of Lorentz invariance means we treat time as distinct from space. 

\medskip
In section 2 we introduce the model, give our motivation and explain how it is constructed. In section 3 we study the Friedman-Robertson-Walker (FRW) universe which results from our model. We close our paper with conclusions and discussions in section 4.

\section{The Model}\label{model}

We propose the following alteration of Einstein's gravity defined by the action
\begin{align}
S=-{3\lambda\over 4\pi G_N^2}\int d^4x\sqrt{-g}\biggl[\sqrt{1+{G_N\over 6\lambda}\left(R+{\cal K}\right)}-q\biggr]\label{DBIgravity} \ ,
\end{align}
plus radiation and matter actions $S_{\rm rad}+S_{\rm M}$.
We work with units  $c=\hbar=1$ and thus the Planck length is $\ell_P=\sqrt{G_N}$.
There are two dimensionless constants $\lambda$ and $q$ introduced. As we will see later, $\lambda$ and $q$ set the scale of inflation and the cosmological constant (CC) at late times, respectively. The scalar quantity ${\cal K}$ is defined by
\be
{\cal K}:=-{2\over \sqrt{h}}\mathsterling_n \left(\sqrt{h}K\right)
=-2\left(K^2+n^{\sigma}\del_{\sigma}K\right)\ ,\label{curlyK}
\ee
where $h_{\mu\nu}\equiv g_{\mu\nu}+n_{\mu}n_{\nu}$ and $K_{\mu\nu}\equiv\half \mathsterling_n h_{\mu\nu}$, with $\mu, \nu=0,1,2,3$, are the induced metric and extrinsic curvature on the space-like surface ${\cal M}$ with the normalised time-like vector ${\bf n}$ normal to ${\cal M}$. 

To be more concrete, we adopt the Arnowitt-Deser-Misner (ADM) decomposition \cite{Arnowitt:1960es}
\be
ds^2=-N^2 dt^2+h_{ij}\left(dx^i+N^i dt\right)\left(dx^j+N^j dt\right)\ ,
\ee
where $N$ and $N_i$, respectively, are the lapse and shift functions, and $h_{ij}$ with $i, j=1, 2, 3$ is the spatial metric which coincides with the induced metric defined above.  
The normal derivative is expressed as
\be
\mathsterling_n=N^{-1}\left(\del_t-\pounds_N\right) \ ,
\ee
with the Lie derivative $\pounds_N$ for the vector field $N_i$. The extrinsic curvature then yields
\be
K_{ij}={1\over 2N}\left(\dot{h}_{ij}-\nabla_iN_j-\nabla_jN_i\right)\ ,
\ee
where $\nabla_i$ is the covariant derivative with respect to the spatial metric $h_{ij}$.

\medskip
A few remarks are in order: 
First, the action (\ref{DBIgravity}) is related to the DBI action, as we shall see shortly, but it is different from Born-Infeld generalisations of Einstein's gravity discussed by Deser and Gibbons \cite{Deser:1998rj}. 
Second, it is noteworthy that the theory defined by the action (\ref{DBIgravity}), though generally covariant, breaks Lorentz invariance by introducing the preferred time-like vector ${\bf  n}$. 
However, in weakly curved space-times, {\it i.e.}, $G_N(R+{\cal K})\ll \lambda$, it reduces to the Einstein-Hilbert action with the (space-like) Gibbons-Hawking-York boundary term  \cite{Gibbons:1976ue, York:1972sj}.

\medskip
The motivation to consider the action (\ref{DBIgravity}) is based on the following observation: with the conformally flat ansatz
\be
ds^2=\ell_P^2\,\phi(\tau,\vec{x})^2\left[-d\tau^2+d\vec{x}^2\right]\ ,\label{ansatz}
\ee
the Einstein-Hibert action (with a cosmological constant) yields the $\lambda\phi^4$ theory
\begin{align}
S_{\rm EH}=-{1\over 16\pi G_N}\int d^4x\sqrt{-g}(R+\lambda)
=-{3\over 4\pi }\int d^4x\left[\half\eta^{\mu\nu}\del_{\mu}\phi\del_{\nu}\phi+{1\over 4\cdot 3}G_N\lambda\phi^4\right] \ ,
\end{align}
up to total derivative terms.
As advocated by Polyakov \cite{Polyakov:2006bz}, the IR triviality of the $\lambda\phi^4$ theory might play a role in explaining the smallness of the CC.\footnote{Assuming an appropriate Wick-rotation} We would rather give significance, however, to the fact that this effective scalar theory is classically conformal. In fact there is a {\it unique} generalisation of the (classical) conformal scalar theory \cite{Maldacena:1997re}:
\be
S_{\rm CFT}=-T_{\rm D3}\int d^4x \lambda\phi^{4}\left(\sqrt{1+{\eta^{\mu\nu}\del_{\mu}\phi\del_{\nu}\phi\over \lambda\phi^{4}}}-q\right)\ ,\label{CFT}
\ee
where $\lambda$ and $q$ are the two arbitrary constants which cannot be fixed by the conformal invariance and will be identified with those in (\ref{DBIgravity}). For a large value of $\phi$, this reduces to the $\lambda\phi^4$ theory.
Note that in type IIB superstring theory on $AdS_5\times S^5$, this is the DBI action for a D3-brane located in $AdS_{5}$ with $\phi$ as the AdS radial coordinate in the Poincar\'e patch, where the constant $T_{\rm D3}$ denotes the D3-brane tension. In the dual ${\cal N}=4$ super Yang-Mills theory, this theory appears as the effective theory for the modulus field $\phi\equiv\sqrt{\Phi^i\Phi^i}$ of the adjoint scalars in the Bogomolny-Prasad-Sommerfeld (BPS) saturated case $q=1$ \cite{Jevicki:1998ub}.
Incidentally, this is the same action as the one DBI inflation is based on \cite{Alishahiha:2004eh}. Note, however, that the scalar field $\phi$ is not an inflaton but the scale factor of the metric in our case.

\medskip
We first note that, due to the classical scale invariance, the $\lambda\phi^4$ theory has the solution $\phi=\pm 1/(\sqrt{\lambda}\tau)$. Plugging this into (\ref{ansatz}), one obtains de Sitter space.
Likewise, the scale invariance of the conformal scalar theory (\ref{CFT}) guarantees the existence of the solution $\phi\propto 1/\tau$. As we will see below, the conformal scalar behaves $\phi\sim 1/(\sqrt{\lambda}\tau)$, near $\phi=0$, and $\phi\sim 1/(\sqrt{\lambda(1-q^{-2})}\tau)$, near $\phi=\infty$, in the generic case. 
This has an interesting cosmological interpretation, provided we alter the Einstein-Hilbert action so that it yields the conformal scalar theory (\ref{CFT}) for the conformally flat universe.
Namely, the universe will have two stages of de Sitter expansion with the CC $E_{\rm inf}^2\equiv\lambda\ell_P^{-2}$ at early times (inflation) and $E_{\Lambda}^2\equiv\lambda(1-q^{-2})\ell_P^{-2}$ at late times (present acceleration). Choosing $q=1+\epsilon$ with $0<\epsilon\ll 1$, we can realise a large hierarchy ${\cal O}(10^{27})$ between the two CC's. We shall discuss this apparent fine tuning in Sec. 4. Upon the inclusion of radiation and matter, the model results in inflation at early times, followed by radiation- and matter-dominated epochs and a subsequent acceleration at late times. 

\medskip
The model (\ref{DBIgravity}) is designed so that it reduces to the conformal scalar theory (\ref{CFT}), when the universe is homogenous and conformally flat, namely $\phi=\phi(\tau)$.\footnote{Observe, however, that this is not so for the generic inhomogeneous conformally flat ansatz.} To see it, note first that the derivative term $(\del\phi)^2/\phi^4$ inside the square root in (\ref{CFT}) is almost the scalar curvature, since 
\be
R=-6\frac{\phi\,\eta^{\mu\nu}\del_{\mu}\del_{\nu}\phi}{\phi^4} \ \longrightarrow \ 6\frac{\ddot{\phi}}{\phi^3} \ ,
\ee 
with the ansatz (\ref{ansatz}) and where the `arrow' specializes to the homogeneous ansatz.  Thus, we need to convert the second derivative of $\phi$ into a first derivative. Herein, we convert only the time derivative, at the cost of giving up on Lorentz invariance.
It is then straightforward to check that (\ref{curlyK}) yields the desired result\footnote{Indeed $\mathcal{K}=-6(\phi^{-3}\ddot{\phi}+\phi^{-4}\dot{\phi}^2)$.}, as anticipated from the role played by the Gibbons-Hawking-York term.

Let us conclude this section by mentioning a more generic class of models that obeys the property discussed in the last paragraph, which has motivated our proposal.\footnote{We thank the anonymous referee of our paper for raising the possibility of this generalization.} Consider the theory
\begin{align}
S=-{3\lambda\over 4\pi G_N^2}\int d^4x\sqrt{-g}\biggl[\sqrt{1+{G_N\over 6\lambda}\left(R+{\cal K} +\alpha\nabla^\mu a_\mu+\beta a_\mu a^\mu+\dots\right)}-q\biggr]\label{DBIgravitya} \ ,
\end{align}
where $a^\nu=n^\mu\nabla_\mu n^\nu$ is the acceleration of the unit time-like vector ${\bf n}$; $\alpha$, $\beta$ are arbitrary coefficients and the dots represent operators of the same type but manifestly non-covariant, such as $\nabla^i a_i$ or with higher spatial derivatives. Since, using the obvious ADM variables for (\ref{ansatz}), we have $a^{\mu}=(0,-\delta^{ij}\partial_j\phi/\phi^3)$, then
\be
\nabla^\mu a_\mu=-\frac{\partial_i \partial^i{\phi}}{\phi^3}-\frac{\partial_i{\phi}\partial^i\phi}{\phi^4} \ \longrightarrow \ 0 \ ,
\ee
\be
a^\mu a_\mu=\frac{\partial_i{\phi}\partial^i\phi}{\phi^4}  \ \longrightarrow \ 0 \ ;
\ee
the addition of these terms is still consistent with our guiding principle. We observe, however, that if $\beta\neq 0$, the model does not resemble Einstein's gravity with a Gibbons-Hawking-York boundary term in weakly curved space-times. Thus we take $\beta=0$. Taking $\alpha\neq 0$ on the other hand, yields a potentially interesting model. Indeed, this term is a total derivative and therefore does not spoil the desired form of the model in weakly curved spacetime. Moreover, adding this term with an appropriate coefficient can yield a Lorentz invariant scalar field model  for the ansatz (\ref{ansatz}) (but does not restore space-time Lorentz invariance). We shall not pursue this more generic class of models further herein. It would be interesting, nevertheless, to see if taking $\alpha\neq 0$ changes the space of solutions of $n$-DBI gravity found in \cite{Herdeiro:2011im}. 

Finally, we do not wish to break the covariance of the model explicitly. In other words, the manifestly non-covariant terms, such as  $\nabla^i a_i$ or with higher spatial derivatives are excluded from the outset.

%
%
%

\section{FRW universe}\label{FRW}
To analyse the advertised remarkable consequences for cosmology of the model (\ref{DBIgravity}), we start by considering the flat FRW model in the conformal frame, {\it i.e.}, (\ref{ansatz}) with $\phi=\phi(\tau)$.
Then, as emphasised, our model (\ref{DBIgravity}) reduces to the conformal scalar theory (\ref{CFT}) in the mini-superspace
\be
S=-{3\over 4\pi}\int d^3x\,d\tau\,\lambda\phi^4\left(\sqrt{1-{\dot{\phi}^2\over\lambda\phi^4}}-q\right)\ .\label{CFTmini}
\ee
The classical dynamics of the scalar field $\phi$ can be understood intuitively from the D3-brane perspective. The second term with $q>0$ corresponds to the repulsive force due to the Ramond-Ramond 5-form flux. When $q=1$, the repulsive force balances the attractive force in the Anti-de-Sitter (AdS) space coming from the first term. This implies that when $q>1$, the repulsive force dominates. So the D3-brane initially located at the AdS horizon $\phi=0$ is repelled to the AdS boundary $\phi=\infty$.

\medskip
Let us see it more in detail. 
Since energy is conserved in the field theory (\ref{CFTmini}), the equation of motion is equivalent to
\be
\lambda\phi^{4}\left({1\over\sqrt{1-{\dot{\phi}^2\over \lambda\phi^{4}}}}
-q\right)={\cal E}\ ,\label{EOM1}
\ee
where ${\cal E}$ is the energy density of this system. Note that from the viewpoint of the scalar field dynamics, ${\cal E}$ is just an integration constant. From the gravity viewpoint, however, this is interpreted as the radiation energy. In other words, we need to add the radiation action $S_{\rm rad}$ to the gravity action (\ref{DBIgravity}) to consider ${\cal E}\ne 0$. This point can be clarified by taking the Einstein gravity limit, $\lambda\to\infty$ and $q\to 1$ with $\lambda(1-q)$ fixed finite, and rewriting (\ref{EOM1}) in terms of the scale factor $a(t)$. This yields the Friedmann equation  (\ref{expand}) with $k={\cal E}_{\rm M}=0$ and the replacement $\lambda(1-q^{-2})$ by $\lambda(q-1)$. It then is clear that ${\cal E}$ corresponds to the radiation energy.

\medskip
It is illustrative to rewrite (\ref{EOM1}) as
\be
\half\dot{\phi}^2+V(\phi)=0 \qquad\mbox{with}\qquad
V(\phi)=-\half\lambda\phi^4\left[1-\left(q+{{\cal E}\over\lambda\phi^4}\right)^{-2}\right]\ .\label{effectivepot}
\ee

\begin{figure}
\centering
\hspace{-1.cm}
\mbox{\subfigure{\includegraphics[height=1.8in]{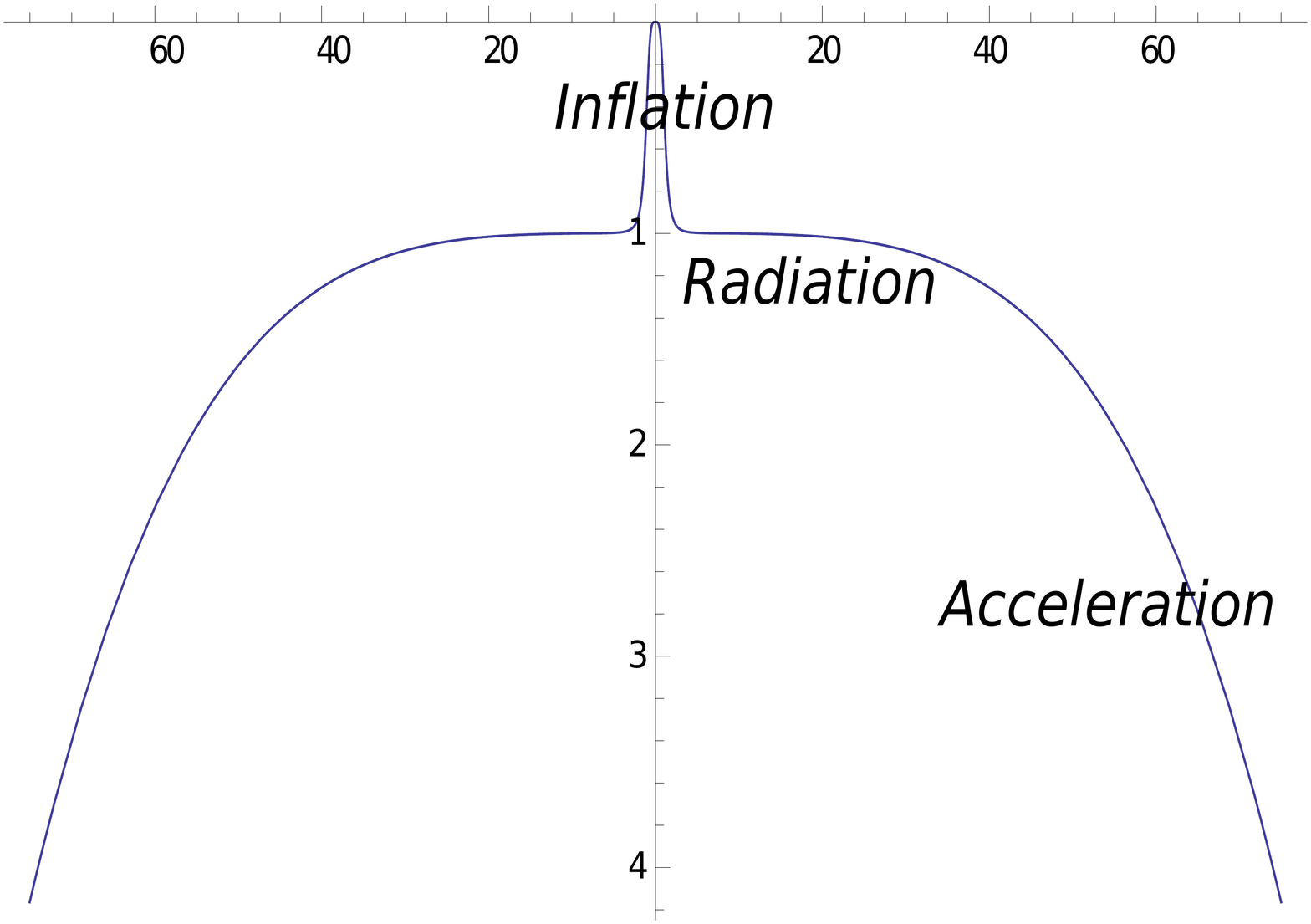}}\hspace{1.3cm}
\subfigure{\includegraphics[height=1.8in]{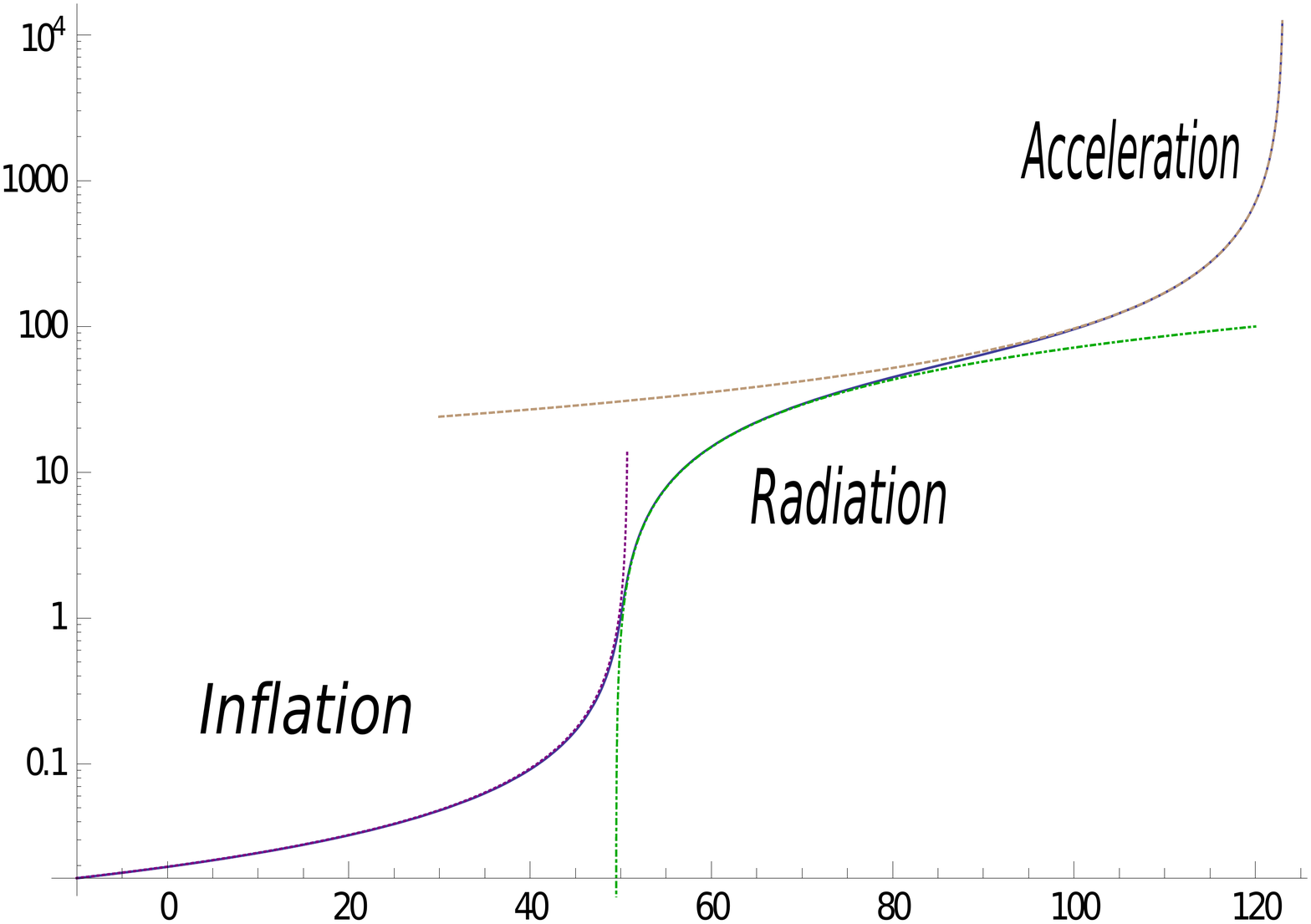} }}
\caption{(Left panel) The effective potential $V(\phi)$ for $q=1+10^{-7}$ with $\lambda={\cal E}=1$. In all Figures we take $\ell_P=1$. (Right panel) The evolution of the universe, $\log\phi$ vs. conformal time $\tau$. There are three distinct stages of evolution; (1) Near de Sitter inflation at early times, (2) Radiation-domination, and (3) Near de Sitter acceleration at late times. In the right panel, the dotted purple, dot-dashed green, and dashed brown lines are the curves (\ref{inflation}), (\ref{radiation}), and (\ref{acceleration}), respectively.} 
\label{pot1}
\end{figure}

\noindent
The effective potential $V(\phi)$ and the solution $\phi(\tau)$ for $q=1+\epsilon$ with $0<\epsilon\ll 1$ are shown in the left and right panels of Fig. \ref{pot1}, respectively.\footnote{The readers should not confuse this potential with the inflaton potential.}
There are three distinct regimes: 
\begin{align}
\hspace{-3.5cm}(1)\,\, \mbox{Near de Sitter inflation}:\quad\sqrt{\lambda}\phi\ll \ell_P^{-1}\quad\mbox{and}\quad -(\tau-\tau_{\rm reheat})\gg \ell_P\ ,\nn
\end{align}
\be
 \sqrt{\lambda}\,\phi\sim -{1\over \tau-\tau_{\rm reheat}}\ .\label{inflation}
\ee
This shows that the inflation scale is $E_{\rm inf}=\sqrt{\lambda}\ell_P^{-1}$, as mentioned earlier.
\begin{align}
\hspace{-1.6cm}(2)\,\, \mbox{Radiation-domination}:\quad 1\ll \sqrt{\lambda}\phi/{\cal E}^{1/4}\ll (1-q^{-2})^{-1/4}\quad\mbox{and}\quad \tau>\tau_{\rm reheat}\ ,\nn
\end{align}
\be
\phi\sim\sqrt{2{\cal E}}(\tau-\tau_{\rm reheat})\ ,\label{radiation}
\ee
where $\tau_{\rm reheat}$ is approximately the time when $\sqrt{\lambda}\phi\sim \ell_P^{-1}$ and should roughly be identified with the time of reheating.
\begin{align}
\hspace{-0.1cm}(3)\,\, \mbox{Near de Sitter acceleration}:\quad \sqrt{\lambda}\phi/{\cal E}^{1/4}\gg (1-q^{-2})^{-1/4}\quad\mbox{and}\quad \tau_{\rm reheat}\ll\tau\lesssim\tau_{\rm end}\ ,\nn
\end{align}
\be
\sqrt{\lambda(1-q^{-2})}\,\phi\sim {1\over\tau-\tau_{\rm end}}\ .\label{acceleration}
\ee
Note that this implies the energy scale of the CC to be 
\be
E_{\Lambda}=\sqrt{\lambda(1-q^{-2})}\,\ell_P^{-1}\ll E_{\rm inf} \ ,
\ee 
much smaller than the scale of inflation. The CC is $10^{-12}$ GeV (which implies the Hubble scale $E_{\Lambda}\sim 10^{-60} M_P$) and the inflation scale is believed to be about $10^{15}$ GeV, close to the GUT scale. Thus in order to generate this hierarchy, we must choose $\epsilon=q-1\sim 10^{-110}$. This clearly requires a fine-tuning (of the same order as in the traditional description).

\medskip
Intriguingly, this shows that our model (\ref{DBIgravity}) results in the flat FRW universe with inflation at early times, followed by the radiation-dominated epoch\footnote{It is not essential that there is only the radiation-dominated epoch at this point. As we will see shortly, matter can be easily added and the matter-domination follows the radiation-dominated period.} and a subsequent acceleration at late times. Both inflation and late time acceleration are nearly de Sitter with a large hierarchy, $1/\sqrt{1-q^{-2}}\sim{\cal O}(\epsilon^{-1/2})$, between the effective cosmological constants. There is no scalar field agent of inflation, and the transition from the inflation to radiation-dominated period is smooth.
Note that the very beginning of the universe at $\tau\to-\infty$ is exactly de Sitter with only a coordinate singularity. (This coordinate patch covers only half of de Sitter space; coordinates may be introduced that extend beyond this singularity  \cite{HE}. This means that, precisely speaking, our flat universe has no beginning of time; it is past eternal.)

\medskip
We also emphasise that in this model gravity becomes approximately Einstein's gravity from the radiation-dominated epoch onwards. At this period, the equation of motion approximates to the Friedmann equation, as will be elaborated below in (\ref{EOM2}) and (\ref{expand}). In other words, the curvature $G_N(R+{\cal K})/(6\lambda)$ in (\ref{DBIgravity}) is much smaller than unity. The deviation from Einstein's gravity becomes significant only in the inflation period. At early times, the curvature $G_N(R+{\cal K})/(6\lambda)\sim -1$ and the square root action is very small. As we will discuss later, this may indicate that the model (\ref{DBIgravity}) predicts a rather large non-Gaussianity in the spectrum of the CMB.

\begin{figure}[h!]
\centering
\includegraphics[height=2.1in]{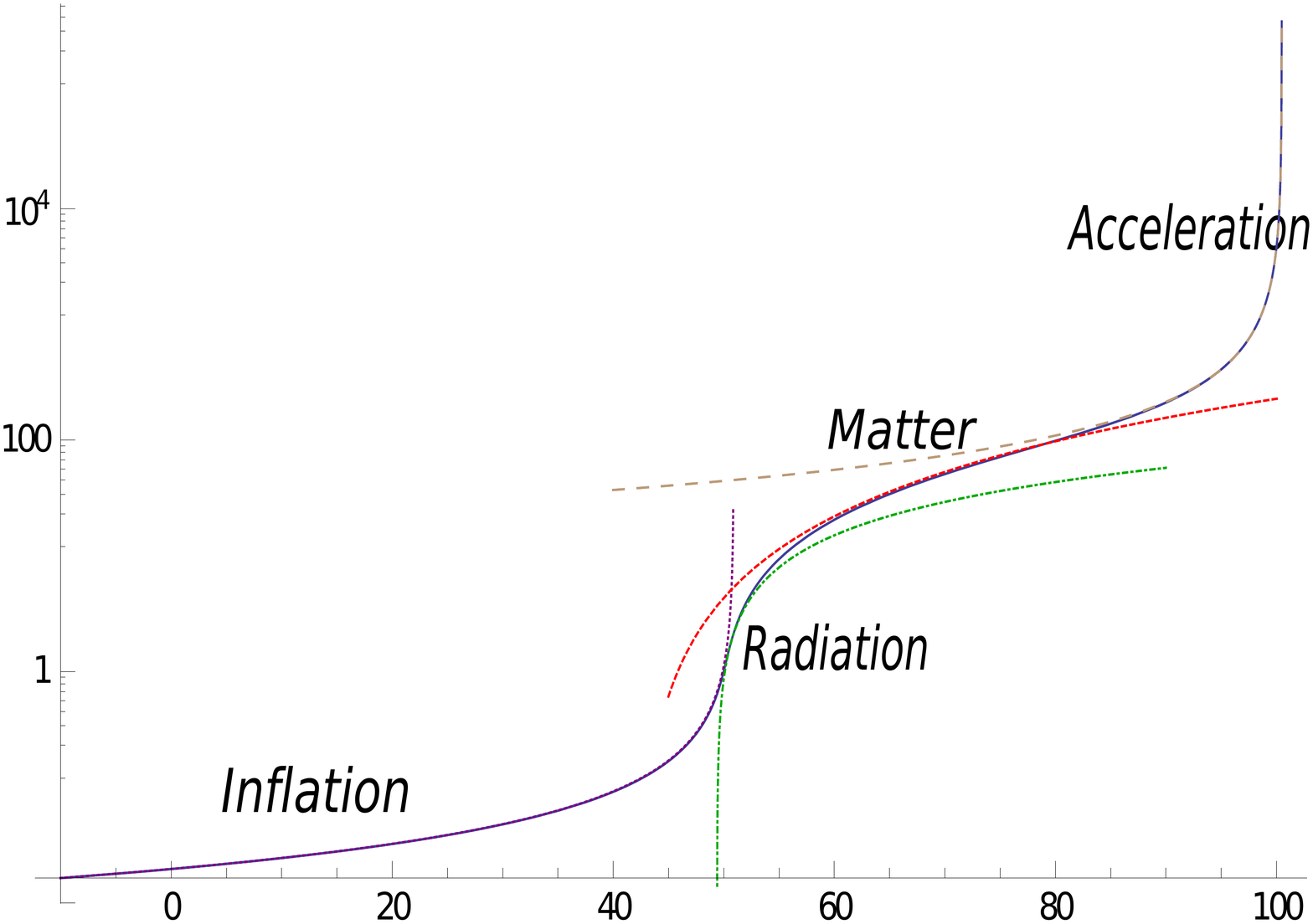}
\caption{The evolution of the flat universe, $\log\phi$ vs. conformal time $\tau$, with matter for $q=1+10^{-7}$ with $\lambda={\cal E}=1$ and ${\cal E}_{\rm M}=10^{-1}$. There are four distinct epochs; (1) Near de Sitter inflation at early times, (2) Radiation-dominated, (3) Matter-dominated, and (4) Near de Sitter late time acceleration. The dotted purple, dot-dashed green, long-dotted red, and dashed brown curves are (\ref{inflation}), (\ref{radiation}), (\ref{matter}), and (\ref{acceleration}), respectively. }
\label{evolve2}
\end{figure}

\medskip
We can easily include matter and the spatial curvature. The FRW universe with spatial curvature $k$ (not normalised to unit) is given by 
\be
ds^2=-dt^2+a(t)^2\left[{dr^2\over 1-kr^2}+r^2(d\theta^2+\sin^2\theta d\varphi^2)\right]\ ,\label{FRW2}
\ee
where the scale factor $a(t)$ and the observer time $t$ are related to the conformal factor $\phi(\tau)$ and the conformal time $\tau$ by
\be
dt=\ell_P\, \phi(\tau)d\tau\qquad\qquad\mbox{and}\qquad\qquad
a(t)=\ell_P\,\phi(\tau)\ .
\ee
The effect of the curvature $k$ is only reflected in the scalar curvature $R=6\phi^{-4}(\phi\ddot{\phi}+k\phi^2)$.
Matter adds the term ${\cal E}_{\rm M}\phi$ to the RHS of (\ref{EOM1}); this becomes clear once we express the equation of motion in terms of the scale factor $a(t)$, yielding a modified Friedmann equation:
\be
\ell_P^{2}H^2=\lambda\left[1+{k\ell_P^{2}\over \lambda a^2}-{\left(1+{k\ell_P^{2}\over\lambda a^2}\right)^2\over \left(q+{{\cal E}\ell_P^{4}\over\lambda a^4}+{{\cal E}_{\rm M}\ell_P^{3}\over\lambda a^3}\right)^{2}}\right]\ ,\label{EOM2}
\ee
where $H$ is the Hubble parameter $a^{-1}da/dt$. Note that for sufficiently large $a$, this expands to the usual Friedmann equation
\be
\ell_P^{2}H^2\sim \lambda\left(1-q^{-2}\right)-{k\ell_P^{2}\over  a^2}+{2{\cal E}\ell_P^{4}\over a^4}
+{2{\cal E}_{\rm M}\ell_P^{3}\over a^3}\ ,\label{expand}
\ee
where we used $q\sim 1$. It is now clear that we can identify
\be
{8\pi G_N^2\rho_{\rm c}\over 3}\,\Omega_{\rm rad}a_0^4=2{\cal E}\ell_P^{4}\ ,\qquad\qquad\mbox{and}\qquad\qquad
{8\pi G_N^2\rho_{\rm c}\over 3}\,\Omega_{\rm M}a_0^3=2{\cal E}_{\rm M}\ell_P^{3}\ .
\ee
The constant $a_0$ is the scale factor at present and $\rho_{\rm c}\equiv 8\pi G_N/(3H_0^2)$ is the critical density.
In Fig. \ref{evolve2} the evolution of the {\it flat} universe with matter is plotted.
In the matter-dominated period, the evolution is approximately given by
\begin{align}
\hspace{-0.5cm}(4)\,\, \mbox{Matter-domination}:\quad {\cal E}/{\cal E}_{\rm M}, \left({\cal E}_{\rm M}/\lambda\right)^{1/3}\ll\phi\ll \left({\cal E}_{\rm M}/\lambda(1-q^{-2})\right)^{1/3}\!\!\quad\mbox{and}\quad \tau> \tau_{\rm reheat}\ ,\nn
\end{align}
\be
\phi\sim {{\cal E}_{\rm M}\over 2}(\tau-\tau_{\rm M})^2\qquad\Longrightarrow\qquad
a\sim\left(9{\cal E}_{\rm M}\ell_P/2\right)^{1/3}(t-t_{\rm M})^{2/3}\ ,\label{matter}
\ee
where the numerics suggests $\tau_{\rm M} < \tau_{\rm reheat}$. 
As we remarked in the last footnote, the inclusion of matter simply adds the matter-dominated epoch taking over the radiation-dominated period. 

\medskip
Note that comparing (\ref{EOM2}) and (\ref{expand}), it is easy to understand why our model behaves better, when the scale factor is very small,  than Einstein's gravity. For simplicity, let us focus on the flat case $k=0$. In the latter, radiation dominates for sufficiently small $a$, whereas in the former, the CC term responsible for inflation is the leading term.\footnote{Of course, since quantum gravity fluctuations presumably become important beyond the inflation scale, one should not trust the Friedmann equation for very small $a$. For argument's sake, we are extrapolating the Friedmann equation beyond the inflation scale.} This is how the initial singularity is evaded in our model. In the case $k\ne 0$, our model still behaves better than Einstein's gravity, but it is not completely singularity free, as we discuss next.

\begin{figure}
\centering
\hspace{-1.cm}
\mbox{\subfigure{\includegraphics[height=1.7in]{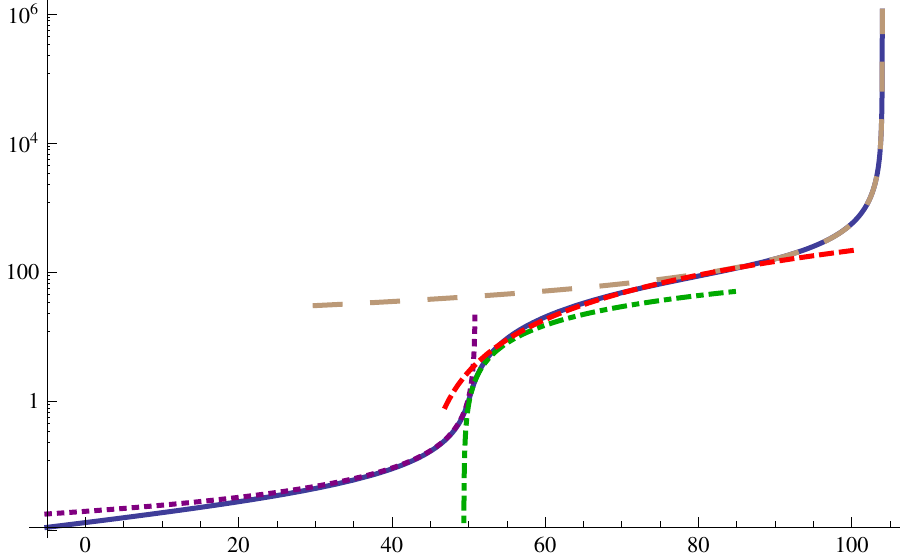}}\hspace{1.5cm}
\subfigure{\includegraphics[height=1.7in]{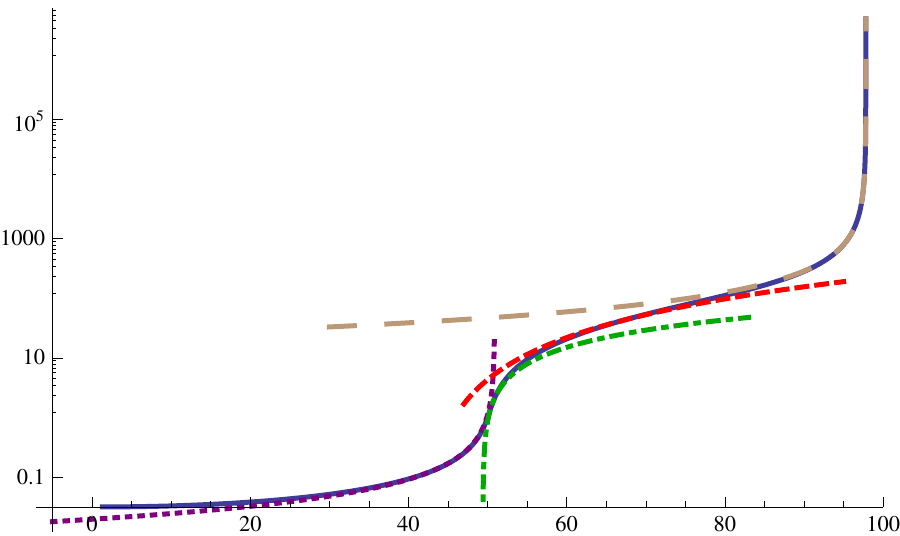} }}
\caption{The evolution of the universe, $\log\phi$ vs. conformal time $\tau$, for $q=1+10^{-7}$ with $\lambda={\cal E}=1$, ${\cal E}_{\rm M}=10^{-1}$, and the spatial curvature $|k|=10^{-3}$. Positive curvature on the left panel and negative curvature on the right panel. At early times, the blue line (actual evolution) deviates a little from the dotted purple line (de Sitter). For the positive curvature, the deviation is downwards, whereas it is upwards for the negative curvature.} 
\label{evolve3}
\end{figure}

\begin{figure}
\centering
\hspace{-1.cm}
\mbox{\subfigure{\includegraphics[height=1.5in]{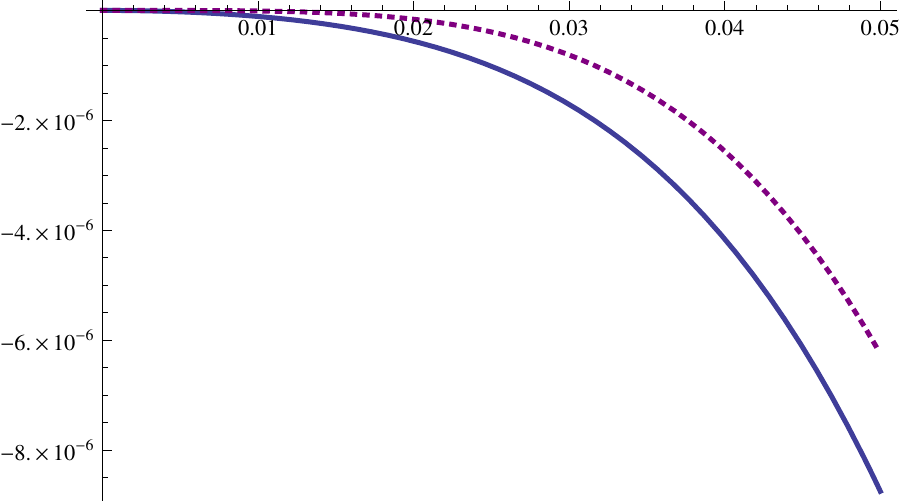}}\hspace{1.5cm}
\subfigure{\includegraphics[height=1.5in]{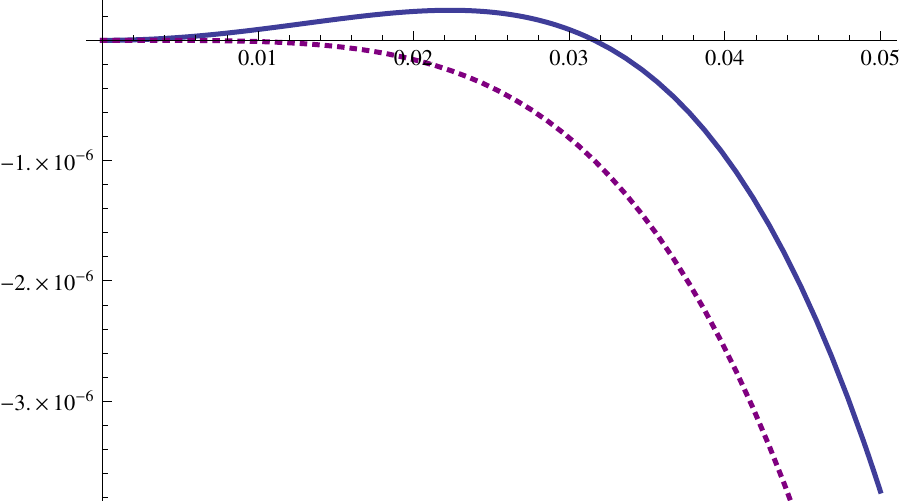} }}
\caption{The effective potential $V(\phi)$ (blue line) near $\phi=0$ for $q=1+10^{-7}$ with $\lambda={\cal E}=1$, ${\cal E}_{\rm M}=10^{-1}$, and the spatial curvature $|k|=10^{-3}$. Positive curvature on the left panel and negative curvature on the right panel. The dotted purple line is the effective potential for $k=0$. These show the early time deviations from the flat universe evolution.} 
\label{pot2}
\end{figure}

\medskip
In open and closed universes, 
the spatial curvature, if not too large, does not change much of the evolution except for the early stage of inflation. The evolution of the universe with the spatial curvature is shown in Fig. \ref{evolve3}, where a slight deviation from near de Sitter inflation is visible in earlier stages of inflation. At early times, the scale factor $a\ll 1/\sqrt{\lambda}$ and thus the equation (\ref{EOM2}) approximates to
\be
\ell_P^{2}H^2\simeq\lambda\left(1+{k\ell_P^{2}\over \lambda a^2}\right)\ .\label{EOMearly}
\ee
The corresponding effective potentials $V(\phi)$ are plotted in Fig. \ref{pot2}. Note that 
this is very different from the standard case (\ref{expand}) where the curvature contribution is negligible at early times, comparing to the radiation and matter. So one might wonder if inflation still solves the flatness problem in this model. 

\medskip
The first term drives near de Sitter inflation with the scale $E_{\rm inf}=\sqrt{\lambda}\ell_P^{-1}$ and the curvature contribution always becomes significant for sufficiently small $a$. The positive and negative $k$'s yield faster and slower evolutions, respectively, relative to the flat universe. This explains the downwards and upwards shifts of evolution for positive and negative $k$'s, respectively, in Fig. \ref{evolve3}. In order for the inflation to occur at all, the curvature scale must be well smaller than the inflation scale, $k\ll E_{\rm inf}^2=\lambda\ell_P^{-2}$. This is the condition that the first term dominates over the second term in (\ref{EOMearly}) for an appreciable range of $a<1/\sqrt{\lambda}$ to generate the inflation with sufficient $e$-foldings. Note that the same condition is necessary anyway in the standard case too. So, inflation in this model solves the flatness problem in the usual manner.

\medskip
A few remarks are in order: (1) in the closed universe, in contrast to the flat universe, the curvature singularity develops in the infinite past $\tau\to-\infty$. There the curvature contribution dominates and the conformal factor goes like $\phi(\tau)\sim \ell_P^{-1}\exp(\sqrt{k}\tau)$ which yields $a(t)\sim \sqrt{k}\,t$. Of course, there is no reason to trust this model at the Planck scale and beyond. So this only signifies that the absence of singularity, which happens to be the case in the flat universe, is not a universal prediction of our model. (2) In the case of an open universe, formally speaking, bubbles nucleate from \lq\lq nothing", since $\phi=0$ is a meta-stable vacuum, as one can see from the right panel of Fig. \ref{pot2}. The bubbles of true vacuum inflate at early times and then make a smooth transition to the radiation-dominated phase, evading the graceful exit problem. Although it is hard to make sense of \lq\lq nothing", it is noteworthy that this has a certain bearing on the \lq\lq old inflation" of Alan Guth \cite{Guth:1980zm}.

\section{Conclusions and Discussion}\label{conclusion}

We proposed a new model of inflation by altering Einstein's gravity, inspired by the guiding principle of scale invariance. With radiation and matter, this model results in near de Sitter inflation, followed by radiation- and matter-dominated epochs and a subsequent late time near de Sitter acceleration. The transition from inflation to radiation is smooth, and it can achieve a large hierarchy between the two effective CC's. Our model breaks Lorentz invariance, in the gravitational sector, by the introduction of a preferred time-like vector. In weakly curved space-times, however, it resembles Einstein's gravity with the Gibbons-Hawking-York boundary term.\footnote{Observe that even in this limit the symmetry group is different from the full diffeomorphisms group; in that sense, our model resembles but is not exactly identical to Einstein's gravity.} Intriguingly, the Big-Bang singularity is absent in the flat universe\footnote{Even though the energy density does diverge as $a\rightarrow 0$. This makes the description of the matter content in terms of a perfect fluid unreliable; it is still noteworthy, nevertheless, that a diverging energy density does not produce a singular geometry. This is analogous, in spirit, to the fact that a point-like charge in Born-Infeld electrodynamics, produces a regular electric field at its location.}  and the open universe realises \lq\lq old inflation" without  the graceful exit problem.
In a related matter, it should be noted that our model does not appear to have an initial conditions problem.  

\medskip
Similarly to \cite{Horava:2009uw}, since our model is endowed with a foliation by space-like surfaces, a scalar graviton may remain as a physical degree of freedom \cite{Blas:2009yd}. It might be that this scalar mode is, in effect, playing the role of inflaton in our model. The St\"uckelberg formalism used in  \cite{Blas:2009yd} might be of help to elucidate this point.

\medskip

The potential existence of such scalar graviton raises the question if pathologies, of the sort discussed in \cite{Blas:2009yd} in the context of Horava gravity, are present in our model. To address this issue, we note that the action (\ref{DBIgravity}), expressing $R+{\cal K}$ in terms of ADM quantities, $h_{ij},N^i,N$, as explicitly done in \cite{Herdeiro:2011im}, can be rewritten as:
\begin{align}
S=-{3\lambda\over 4\pi G_N^2}\int dtd^3x\sqrt{h}N\biggl[\sqrt{1+{G_N\over 6\lambda}\left(R_3+K_{ij}K^{ij}-K^2-2N^{-1}\Delta N\right)}-q\biggr]\label{DBIgravity20} \ ,
\end{align}
where $R_3$ is the Ricci scalar of $h_{ij}$. In particular, observe the existence of the crucial new term with the lapse $N^{-1}\Delta N$. To understand the importance of this new lapse dependent term, we note that in \cite{Blas:2009qj}, a generic action of type
\be
S=\frac{M_P^2}{2}\int d^3xdt\sqrt{h}N\left(K_{ij}K^{ij}-\lambda K^2 -\mathcal{V}\right) \ ,
\ee
is considered. The basic observation in \cite{Blas:2009qj} is that if the ``potential'' $\mathcal{V}$ depends on both quantities constructed from $h_{ij}$ and on quantities constructed from ${\partial_i N}/{N}$, then the short distance instabilities of the original Horava model are not necessarily present. An analysis (to be presented elsewhere) indeed shows  that no evidence for such instabilities is found in our model.

\medskip
From the radiation-dominated period onwards, our model is no different from the concordance cosmology. However, since it does differ in the inflation period, it may leave distinct signatures on the CMB spectra.   
Due to its near scale invariance in the inflation period, the power spectrum in our model will be nearly scale invariant in consistency with the Harrison-Zel'dovich-Peebles spectrum at low angular momentum $l\lesssim 50$. It remains to be seen, however, whether it can predict a slight red tilt of the spectral index $n_s=0.968\pm 0.012$ \cite{Komatsu:2010fb}.
Also, as mentioned earlier, our model may predict a rather large non-Gaussianity \cite{Maldacena:2002vr}.  Indeed, in the inflation epoch, our model deviates significantly from Einstein's gravity, where higher derivative corrections become important. As remarked, the DBI square root action is small during inflation, so the enhancement of non-Gaussianity may be expected similarly to \cite{Alishahiha:2004eh}.


\section*{Acknowledgement}

We would like to thank the anonymous referee of our paper for various relevant comments. S.H. would like to thank Yuki Sato for useful discussions.
This work was partially supported by the Grant-in-Aid for Nagoya
University Global COE Program (G07) and by FCT (Portugal) through the project CERN/FP/116341/2010.




\begin{thebibliography}{40}

\bibitem{Guth:1980zm}
  A.~H.~Guth,
  ``The Inflationary Universe: A Possible Solution to the Horizon and Flatness Problems,''
  Phys.\ Rev.\  {\bf D23}, 347-356 (1981).

\bibitem{Linde:1981mu}
  A.~D.~Linde,
  ``A New Inflationary Universe Scenario: A Possible Solution of the Horizon,
  Flatness, Homogeneity, Isotropy and Primordial Monopole Problems,''
  Phys.\ Lett.\  B {\bf 108}, 389 (1982);
  ``Chaotic Inflation,''
  Phys.\ Lett.\  B {\bf 129} (1983) 177.

\bibitem{Albrecht:1982wi}
  A.~Albrecht and P.~J.~Steinhardt,
  ``Cosmology for Grand Unified Theories with Radiatively Induced Symmetry
  Breaking,''
  Phys.\ Rev.\ Lett.\  {\bf 48}, 1220 (1982).

\bibitem{Starobinsky:1980te}
  A.~A.~Starobinsky,
  ``A New Type of Isotropic Cosmological Models Without Singularity,''
  Phys.\ Lett.\  {\bf B91}, 99-102 (1980);
  ``Relict Gravitation Radiation Spectrum and Initial State of the Universe. (In Russian),''
  JETP Lett.\  {\bf 30}, 682-685 (1979).

\bibitem{Sato:1980yn}
  K.~Sato,
  ``First Order Phase Transition of a Vacuum and Expansion of the Universe,''
  Mon.\ Not.\ Roy.\ Astron.\ Soc.\  {\bf 195}, 467 (1981).

\bibitem{Lyth:1998xn}
  D.~H.~Lyth, A.~Riotto,
  ``Particle physics models of inflation and the cosmological density perturbation,''
  Phys.\ Rept.\  {\bf 314}, 1-146 (1999).
  [hep-ph/9807278].

\bibitem{Linde:2007fr}
  A.~D.~Linde,
  ``Inflationary Cosmology,''
  Lect.\ Notes Phys.\  {\bf 738}, 1 (2008)
  [arXiv:0705.0164 [hep-th]].


\bibitem{Rubakov:2010hq}
  V.~Rubakov, A.~Vlasov,
  ``What do we learn from CMB observations,''
    [arXiv:1008.1704 [astro-ph.CO]].

\bibitem{Baumann:2008aq}
  D.~Baumann {\it et al.}  [CMBPol Study Team Collaboration],
  ``CMBPol Mission Concept Study: Probing Inflation with CMB Polarization,''
  AIP Conf.\ Proc.\  {\bf 1141}, 10 (2009)
  [arXiv:0811.3919 [astro-ph]].

\bibitem{Komatsu:2010fb}
  E.~Komatsu {\it et al.} [ WMAP Collaboration ],
  Astrophys.\ J.\ Suppl.\  {\bf 192}, 18 (2011).
  [arXiv:1001.4538 [astro-ph.CO]].

\bibitem{HenryTye:2006uv}
  S.~H.~Henry Tye,
  ``Brane inflation: String theory viewed from the cosmos,''
  Lect.\ Notes Phys.\  {\bf 737}, 949 (2008)
  [arXiv:hep-th/0610221].



\bibitem{Kallosh:2007ig}
  R.~Kallosh,
  ``On inflation in string theory,''
  Lect.\ Notes Phys.\  {\bf 738}, 119 (2008)
  [arXiv:hep-th/0702059].

\bibitem{Burgess:2007pz}
  C.~P.~Burgess,
  ``Lectures on Cosmic Inflation and its Potential Stringy Realizations,''
  Class.\ Quant.\ Grav.\  {\bf 24}, S795 (2007).
  [arXiv:0708.2865 [hep-th]].


\bibitem{McAllister:2007bg}
  L.~McAllister and E.~Silverstein,
  ``String Cosmology: A Review,''
  Gen.\ Rel.\ Grav.\  {\bf 40}, 565 (2008)
  [arXiv:0710.2951 [hep-th]].

\bibitem{Jacobson:2008aj}
  T.~Jacobson,
  ``Einstein-aether gravity: A Status report,''
  PoS {\bf QG-PH}, 020 (2007)
  [arXiv:0801.1547 [gr-qc]];
  C.~Eling, T.~Jacobson,
  ``Static postNewtonian equivalence of GR and gravity with a dynamical preferred frame,''
  Phys.\ Rev.\  {\bf D69}, 064005 (2004).
  [gr-qc/0310044].

\bibitem{Horava:2009uw}
  P.~Horava,
  ``Quantum Gravity at a Lifshitz Point,''
  Phys.\ Rev.\  D {\bf 79}, 084008 (2009)
  [arXiv:0901.3775 [hep-th]].

\bibitem{Arnowitt:1960es}
  R.~L.~Arnowitt, S.~Deser and C.~W.~Misner,
  ``Canonical variables for general relativity,''
  Phys.\ Rev.\  {\bf 117}, 1595 (1960);
  R.~L.~Arnowitt, S.~Deser and C.~W.~Misner,
  ``The Dynamics of general relativity,''
  arXiv:gr-qc/0405109.


\bibitem{Deser:1998rj}
  S.~Deser and G.~W.~Gibbons,
  ``Born-Infeld-Einstein actions?,''
  Class.\ Quant.\ Grav.\  {\bf 15}, L35 (1998)
  [arXiv:hep-th/9803049].

\bibitem{Gibbons:1976ue}
  G.~W.~Gibbons and S.~W.~Hawking,
  ``Action Integrals and Partition Functions in Quantum Gravity,''
  Phys.\ Rev.\  D {\bf 15}, 2752 (1977).

\bibitem{York:1972sj}
  J.~W.~York,
  ``Role of conformal three geometry in the dynamics of gravitation,''
  Phys.\ Rev.\ Lett.\  {\bf 28}, 1082 (1972).


\bibitem{Polyakov:2006bz}
  A.~M.~Polyakov,
  ``Beyond space-time,''
  arXiv:hep-th/0602011.

\bibitem{Maldacena:1997re}
  J.~M.~Maldacena,
  ``The Large N limit of superconformal field theories and supergravity,''
  Adv.\ Theor.\ Math.\ Phys.\  {\bf 2}, 231 (1998)
  [Int.\ J.\ Theor.\ Phys.\  {\bf 38}, 1113 (1999)]
  [arXiv:hep-th/9711200].

\bibitem{Jevicki:1998ub}
  A.~Jevicki, Y.~Kazama and T.~Yoneya,
  ``Generalized conformal symmetry in D-brane matrix models,''
  Phys.\ Rev.\  D {\bf 59}, 066001 (1999)
  [arXiv:hep-th/9810146].

\bibitem{Alishahiha:2004eh}
  M.~Alishahiha, E.~Silverstein, D.~Tong,
  ``DBI in the sky,''
  Phys.\ Rev.\  {\bf D70}, 123505 (2004).
  [hep-th/0404084].

\bibitem{HE}
S.~W.~Hawking and G.~F.~R.~Ellis,
``The Large Scale Structure of Space-Time,"
Cambridge University Press.

\bibitem{Blas:2009yd}
  D.~Blas, O.~Pujolas and S.~Sibiryakov,
  ``On the Extra Mode and Inconsistency of Horava Gravity,''
  JHEP {\bf 0910}, 029 (2009)
  [arXiv:0906.3046 [hep-th]].

\bibitem{Herdeiro:2011im}
  C.~Herdeiro, S.~Hirano and Y.~Sato,
  ``n-DBI gravity,''
  Phys.\ Rev.\ D {\bf 84} (2011) 124048
  [arXiv:1110.0832 [gr-qc]].
  
\bibitem{Blas:2009qj}
  D.~Blas, O.~Pujolas and S.~Sibiryakov,
  ``Consistent Extension of Horava Gravity,''
  Phys.\ Rev.\ Lett.\  {\bf 104} (2010) 181302
  [arXiv:0909.3525 [hep-th]].


\bibitem{Maldacena:2002vr}
  J.~M.~Maldacena,
  ``Non-Gaussian features of primordial fluctuations in single field
  inflationary models,''
  JHEP {\bf 0305}, 013 (2003)
  [arXiv:astro-ph/0210603].



\end{thebibliography}
\end{document}